\documentstyle[11pt]{article}
\def\text{}
\renewcommand{\baselinestretch}{1.75}

\def\ha{ {\textstyle {1\over 2}} }
\def\half{ {\textstyle {1\over 2}} }
\def\prodN{ \prod_{i=1}^N }

\def\quarS{{\bar S}}
\newcommand{\be}{\begin{eqnarray}}
\newcommand{\ee}{\end{eqnarray}}

\newbox\bigstrutbox
\setbox\bigstrutbox=\hbox{\vrule height12pt depth5.5pt width0pt}

\def\bigstrut{\relax\ifmmode\copy\bigstrutbox\else\unhcopy\bigstrutbox\fi}

\newcommand{\np}{\newpage}

\newcommand{\vs}{\vspace}

\newcommand{\nn}{\nonumber}

\def\so#1{{\rm SO}( #1)}
\def\su#1{{\rm SU}( #1)}
\def\sp#1{{\rm Sp}( #1)}

\def\la#1{\label{#1}}
\def\eqs#1{(\ref{#1})}

\font\upright=cmu10 
\font\cmss=cmss10 at 11pt \font\cmsss=cmss8 at 8pt

\def\inbar{\vrule height1.5ex width.4pt depth0pt}
\def\mininbar{\vrule height.75ex width.3pt depth0pt} 
\def\cc{\relax\,\hbox{$\mininbar\kern-.2em{\hbox{\rm\tiny
C}}$}}
\def\IZ{\relax\ifmmode\mathchoice
{\hbox{\cmss Z\kern-.4em Z}}{\hbox{\cmss Z\kern-.4em Z}}
{\lower.4pt\hbox{\cmsss Z\kern-.4em Z}}
{\lower1.2pt\hbox{\cmsss Z\kern-.4em Z}}\else{\cmss Z\kern-.4em
Z}\fi}
\def\IC{\relax\,\hbox{$\inbar\kern-.3em{\rm C}$}}
\def\IR{\relax{\rm I\kern-.18em R}}
\def\mt{\rlap{\cmss T}\kern 3.0pt{\hbox{{\cmss T}}}}
\def\identity{{\upright\rlap{1}\kern 2.0pt 1}}
\def\sqr#1#2{{\vcenter{\vbox{\hrule height.#2pt
    \hbox{\vrule width.#2pt height#1pt \kern#1pt
    \vrule width.#2pt}
    \hrule height.#2pt}}}}

\newdimen\tableauside\tableauside=1.1ex
\newdimen\tableaurule\tableaurule=0.4pt
\newdimen\tableaustep
\def\phantomhrule#1{\hbox{\vbox to0pt{\hrule height\tableaurule
width#1\vss}}}
\def\phantomvrule#1{\vbox{\hbox to0pt{\vrule width\tableaurule
height#1\hss}}}
\def\sqr{\vbox{%
  \phantomhrule\tableaustep
 
\hbox{\phantomvrule\tableaustep\kern\tableaustep\phantomvrule\tableaustep}%
  \hbox{\vbox{\phantomhrule\tableauside}\kern-\tableaurule}}}
\def\squares#1{\hbox{\count0=#1\noindent\loop\sqr
  \advance\count0 by-1 \ifnum\count0>0\repeat}}
\def\tableau#1{\vcenter{\offinterlineskip
  \tableaustep=\tableauside\advance\tableaustep by-\tableaurule
  \kern\normallineskip\hbox
    {\kern\normallineskip\vbox
      {\gettableau#1 0 }%
     \kern\normallineskip\kern\tableaurule}%
  \kern\normallineskip\kern\tableaurule}}
\def\gettableau#1 {\ifnum#1=0\let\next=\null\else
  \squares{#1}\let\next=\gettableau\fi\next}
\newcommand{\fund}{\tableau{1}}
\newcommand{\Ysymm}{\tableau{2}}
\newcommand{\Yasymm}{\tableau{1 1}}
\begin{document}

\thispagestyle{empty}

\vs*{-25mm}
\begin{flushright}

BRX-TH-475\\[-.15in]
BOW-PH-118\\[-.15in]
HUTP-00/A021\\[-.15in]
\vs{8mm}
\end{flushright}
\setcounter{footnote}{0}

\begin{center}
{\Large{\bf  Seiberg-Witten curves for elliptic models}}
\renewcommand{\baselinestretch}{1}
\small
\normalsize
\vspace{.1in}
\footnote{Based on a talk 
by S. G. Naculich at the 
{\it Workshop on Strings, Duality and Geometry,} \\
\phantom{aaa} 
University of Montreal, March 2000.
}\\

\vspace{.3in}

Isabel P. Ennes\footnote{Research supported 
by the DOE under grant DE--FG02--92ER40706.}$^{,a}$, 
Carlos Lozano\footnotemark[2]$^{,a}$, 
Stephen G. Naculich\footnote{Research supported in part by the 
National Science Foundation under grant no.~PHY94-07194 through the \\
\phantom{aaa}  ITP Scholars Program.}$^{,b}$, 
Howard J. Schnitzer\footnote{Permanent address.}
${}^{\!\!\!,\!\!\!}$
\footnote{Research supported in part
by the DOE under grant DE--FG02--92ER40706.\\
{\tt \phantom{aaa} naculich@bowdoin.edu; 
ennes,lozano,schnitzer@brandeis.edu}\\}$^{,a,c}$\\

\vspace{.2in}

${}^{a,4}$Martin Fisher School of Physics\\
Brandeis University, Waltham, MA 02454

\vspace{.2in}

${}^{b}$Department of Physics\\
Bowdoin College, Brunswick, ME 04011

\vspace{.2in}

${}^{c}$Lyman Laboratory of Physics \\
Harvard University, Cambridge, MA 02138

\vspace{.3in}

{\bf{Abstract}} 
\end{center}
\renewcommand{\baselinestretch}{1.75}
\small
\normalsize
\begin{quotation}
\baselineskip14pt
\noindent  
Four-dimensional ${\cal N}$=2 gauge theories may be obtained from
configurations of D-branes in type IIA string theory.
Unitary gauge theories with two-index representations,
and orthogonal and symplectic gauge theories, are
constructed from configurations containing orientifold planes.
Models with two orientifold planes imply a compact dimension,
and correspond to elliptic models.  Lifting these configurations
to M-theory allows one to derive the Seiberg-Witten curves for
these gauge theories. We describe how the Seiberg-Witten curves,
necessarily of infinite order, are obtained for these elliptic
models.  These curves are used to calculate the instanton
expansion of the prepotential; we explicitly find the
one-instanton prepotential for all the elliptic models considered.
\end{quotation}

\np 

\setcounter{page}{1}
\noindent{\bf 1. ~Introduction}
\renewcommand{\theequation}{1.\arabic{equation}}
\setcounter{equation}{0}
\baselineskip20pt

In the Seiberg-Witten approach to 
four-dimensional ${\cal N}=2$ supersymmetric 
gauge theories \cite{SeibergWitten},
one begins by identifying an algebraic curve 
and meromorphic differential specific to
the gauge group and matter content of the theory.
One then calculates the periods of this differential, 
and integrates the result to obtain the exact low-energy prepotential 
for the gauge theory.
The perturbative and instanton contributions
to the prepotential may then be compared 
with results obtained directly from the 
microscopic Lagrangian of the gauge theory.
General methods were presented in Refs.~\cite{DHokerKricheverPhong}
for computing the prepotential for gauge theories with hyperelliptic curves.
For theories with non-hyperelliptic curves,
a systematic approximation scheme for calculating 
the instanton expansion of the prepotential 
was developed in Refs.~\cite{oneanti}-\cite{twoanti}
and is described in the talk by H. J. Schnitzer at this Workshop
\cite{montrealone}.

M-theory provides a systematic means of deriving Seiberg-Witten 
curves \cite{WittenM}.
One identifies a type IIA brane configuration that gives
rise to the four-dimensional gauge theory of interest,
and then lifts this to an M5 brane configuration; 
the world-volume of the M5 brane contains the Seiberg-Witten curve
as a factor.

Our goal in this talk is to explain how to derive 
the one-instanton prepotentials
for the class of elliptic models with a simple gauge group \cite{elliptic}.
In sect.~2, we describe the construction of type IIA brane configurations
that correspond to these four-dimensional ${\cal N}=2$ gauge theories.
We explain in sect.~3 how the lift to M-theory 
may be used to obtain the Seiberg-Witten curves for these theories.
In sect.~4, the quartic truncation of the SW curve  is used
to obtain explicit expressions for the one-instanton prepotentials 
for each of these theories.

\vspace{.3in}
\noindent{\bf 2. Type IIA brane configurations and 
four-dimensional gauge theory}
\renewcommand{\theequation}{2.\arabic{equation}}
\setcounter{equation}{0}

We begin by briefly reviewing type IIA brane configurations
associated with various four-dimensional ${\cal N}=2$ gauge theories;
see Ref.~\cite{GiveonKutasov} for a review with references.
A typical brane configuration, shown in fig.~1, contains 
a number of NS 5-branes, extended in the 012345 directions,
located at the same point in the 789 directions, 
and having distinct values of $x_6$.
The horizontal direction in the figure corresponds to $x_6$, 
the vertical direction to $v = x_4 + i x_5$, 
with the remaining directions suppressed. 
The NS 5-branes are connected by D4-branes, 
extended in the 01236 directions, but of finite length along $x_6$.

\begin{picture}(430,200)(10,10)

\put(100,50){\line(0,1){150}}
\put(220,50){\line(0,1){150}}
\put(340,50){\line(0,1){150}}



\put(30,120){\vector(0,1){30}}
\put(29,153){$v$}
\put(30,120){\vector(1,0){30}}
\put(64,118){$x_6$}
\put(85,80){$a_1$}
\put(100,80){\line(1,0){9}}
\put(119,80){\line(1,0){9}}
\put(138,80){\line(1,0){9}}
\put(157,80){\line(1,0){9}}
\put(176,80){\line(1,0){9}}
\put(195,80){\line(1,0){9}}
\put(211,80){\line(1,0){9}}

\put(85,100){$a_2$}
\put(100,100){\line(1,0){9}}
\put(119,100){\line(1,0){9}}
\put(138,100){\line(1,0){9}}
\put(157,100){\line(1,0){9}}
\put(176,100){\line(1,0){9}}
\put(195,100){\line(1,0){9}}
\put(211,100){\line(1,0){9}}

\put(80,180){$a_{N_1}$}
\put(100,180){\line(1,0){9}}
\put(119,180){\line(1,0){9}}
\put(138,180){\line(1,0){9}}
\put(157,180){\line(1,0){9}}
\put(176,180){\line(1,0){9}}
\put(195,180){\line(1,0){9}}
\put(211,180){\line(1,0){9}}

\put(155,120){$\cdot$}
\put(155,125){$\cdot$}
\put(155,130){$\cdot$}
\put(155,115){$\cdot$}

\put(220,90){\line(1,0){9}}
\put(239,90){\line(1,0){9}}
\put(258,90){\line(1,0){9}}
\put(277,90){\line(1,0){9}}
\put(296,90){\line(1,0){9}}
\put(315,90){\line(1,0){9}}
\put(331,90){\line(1,0){9}}
\put(350,90){$b_{1}$}

\put(220,105){\line(1,0){9}}
\put(239,105){\line(1,0){9}}
\put(258,105){\line(1,0){9}}
\put(277,105){\line(1,0){9}}
\put(296,105){\line(1,0){9}}
\put(315,105){\line(1,0){9}}
\put(331,105){\line(1,0){9}}
\put(345,105){$b_{2}$}

\put(220,170){\line(1,0){9}}
\put(239,170){\line(1,0){9}}
\put(258,170){\line(1,0){9}}
\put(277,170){\line(1,0){9}}
\put(296,170){\line(1,0){9}}
\put(315,170){\line(1,0){9}}
\put(331,170){\line(1,0){9}}
\put(345,170){$b_{N_2}$}

\put(275,120){$\cdot$}
\put(275,125){$\cdot$}
\put(275,130){$\cdot$}
\put(275,135){$\cdot$}
\put(220,5){\makebox(0,0)[b]{\bf {Figure 1}}}
\end{picture}

The brane configuration in fig.~1 gives rise to an ${\cal N}=2$
$\su{N_1}\times\su{N_2}$ gauge theory with a matter
hypermultiplet in the bifundamental representation \cite{WittenM}.
The first two NS 5-branes are connected by $N_1$ D4-branes;
strings extending between the latter give rise
to the adjoint vector multiplet  of the gauge group $\su{N_1}$.
Strings extending between the
$N_2$ D4-branes connecting the last two NS 5-branes
yield the adjoint vector multiplet of the gauge group $\su{N_2}$.
Finally, 
strings extending between the D4 branes 
connecting the first two NS 5-branes 
and the D4 branes connecting the last two NS 5-branes 
give rise to the hypermultiplet
in the $(\overline{\fund}, \fund)$ representation
of $\su{N_1} \times \su{N_2}$.

\begin{picture}(430,200)(10,10)

\put(100,50){\line(0,1){150}}
\put(220,50){\line(0,1){150}}
\put(340,50){\line(0,1){150}}



\put(30,120){\vector(0,1){30}}
\put(29,153){$v$}
\put(30,120){\vector(1,0){30}}
\put(64,118){$x_6$}

\put(200,124){O$6$}
\put(216,124){$\otimes$}
\put(225,124){$-\half m$}

\put(85,80){$a_1$}
\put(100,80){\line(1,0){9}}
\put(119,80){\line(1,0){9}}
\put(138,80){\line(1,0){9}}
\put(157,80){\line(1,0){9}}
\put(176,80){\line(1,0){9}}
\put(195,80){\line(1,0){9}}
\put(211,80){\line(1,0){9}}

\put(85,100){$a_2$}
\put(100,100){\line(1,0){9}}
\put(119,100){\line(1,0){9}}
\put(138,100){\line(1,0){9}}
\put(157,100){\line(1,0){9}}
\put(176,100){\line(1,0){9}}
\put(195,100){\line(1,0){9}}
\put(211,100){\line(1,0){9}}

\put(80,161){$a_{N}$}
\put(100,161){\line(1,0){9}}
\put(119,161){\line(1,0){9}}
\put(138,161){\line(1,0){9}}
\put(157,161){\line(1,0){9}}
\put(176,161){\line(1,0){9}}
\put(195,161){\line(1,0){9}}
\put(211,161){\line(1,0){9}}

\put(155,120){$\cdot$}
\put(155,125){$\cdot$}
\put(155,130){$\cdot$}
\put(155,115){$\cdot$}

\put(220,90){\line(1,0){9}}
\put(239,90){\line(1,0){9}}
\put(258,90){\line(1,0){9}}
\put(277,90){\line(1,0){9}}
\put(296,90){\line(1,0){9}}
\put(315,90){\line(1,0){9}}
\put(331,90){\line(1,0){9}}
\put(345,90){$-a_N -m$}

\put(220,152){\line(1,0){9}}
\put(239,152){\line(1,0){9}}
\put(258,152){\line(1,0){9}}
\put(277,152){\line(1,0){9}}
\put(296,152){\line(1,0){9}}
\put(315,152){\line(1,0){9}}
\put(331,152){\line(1,0){9}}
\put(345,152){$-a_2 -m$}

\put(220,170){\line(1,0){9}}
\put(239,170){\line(1,0){9}}
\put(258,170){\line(1,0){9}}
\put(277,170){\line(1,0){9}}
\put(296,170){\line(1,0){9}}
\put(315,170){\line(1,0){9}}
\put(331,170){\line(1,0){9}}
\put(345,170){$-a_1 -m$}

\put(275,120){$\cdot$}
\put(275,125){$\cdot$}
\put(275,130){$\cdot$}
\put(275,135){$\cdot$}
\put(220,10){\makebox(0,0)[b]{\bf {Figure 2}}}
\end{picture}

Figure 2 contains an orientifold 6-plane 
extending in the 0123789 directions, 
and intersecting the central NS 5-brane.
The orientifold 6-plane can have either $+4$ or $-4$ units
of 6-brane charge, and is designated O$6^+$ or O$6^-$ respectively.
The O6 plane identifies the points $(x_6, v) \sim (-x_6, -v-m)$  
in the directions transverse to it.
In terms of the gauge theory,
the orientifold identifies the two factors of the gauge group
$\su{N}\times \su{N}$,
and projects the bifundamental representation onto 
either the symmetric $\Ysymm$ representation (O$6^+$)
or the antisymmetric $\Yasymm$ representation (O$6^-$) \cite{LandsteinerLopez}.


\begin{picture}(430,200)(10,10)

\put(181,50){\line(0,1){150}}
\put(320,50){\line(0,1){150}}


\put(90,120){\vector(0,1){30}}
\put(89,153){$v$}
\put(90,120){\vector(1,0){30}}
\put(124,118){$x_6$}


\put(181,80){\line(1,0){9}}
\put(200,80){\line(1,0){9}}
\put(219,80){\line(1,0){9}}
\put(238,80){\line(1,0){9}}
\put(257,80){\line(1,0){9}}
\put(276,80){\line(1,0){9}}
\put(295,80){\line(1,0){9}}
\put(311,80){\line(1,0){9}}

\put(181,100){\line(1,0){9}}
\put(200,100){\line(1,0){9}}
\put(219,100){\line(1,0){9}}
\put(238,100){\line(1,0){9}}
\put(257,100){\line(1,0){9}}
\put(276,100){\line(1,0){9}}
\put(295,100){\line(1,0){9}}
\put(311,100){\line(1,0){9}}

\put(181,140){\line(1,0){9}}
\put(200,140){\line(1,0){9}}
\put(219,140){\line(1,0){9}}
\put(238,140){\line(1,0){9}}
\put(257,140){\line(1,0){9}}
\put(276,140){\line(1,0){9}}

\put(181,80){\line(1,0){9}}
\put(200,80){\line(1,0){9}}
\put(219,80){\line(1,0){9}}
\put(238,80){\line(1,0){9}}
\put(257,80){\line(1,0){9}}
\put(276,80){\line(1,0){9}}
\put(295,80){\line(1,0){9}}
\put(311,80){\line(1,0){9}}

\put(181,100){\line(1,0){9}}
\put(200,100){\line(1,0){9}}
\put(219,100){\line(1,0){9}}
\put(238,100){\line(1,0){9}}
\put(257,100){\line(1,0){9}}
\put(276,100){\line(1,0){9}}
\put(295,100){\line(1,0){9}}
\put(311,100){\line(1,0){9}}

\put(181,140){\line(1,0){9}}
\put(200,140){\line(1,0){9}}
\put(219,140){\line(1,0){9}}
\put(238,140){\line(1,0){9}}
\put(257,140){\line(1,0){9}}
\put(276,140){\line(1,0){9}}
\put(295,140){\line(1,0){9}}
\put(311,140){\line(1,0){9}}

\put(181,160){\line(1,0){9}}
\put(200,160){\line(1,0){9}}
\put(219,160){\line(1,0){9}}
\put(238,160){\line(1,0){9}}
\put(257,160){\line(1,0){9}}
\put(276,160){\line(1,0){9}}
\put(295,160){\line(1,0){9}}
\put(311,160){\line(1,0){9}}

\put(245,120){$\otimes$}

\put(245,10){\makebox(0,0)[b]{\bf {Figure 3}}}

\end{picture}

In fig.~3, the orientifold plane is located midway
between the NS 5-branes,
and serves to project out some of the states of the adjoint
representation of the $\su{2N}$ gauge group,
leaving either $\so{2N}$ for O$6^+$ or $\sp{2N}$ for O$6^-$
\cite{LandsteinerLopez}.

Most asymptotically-free ${\cal N}=2$ gauge theories 
can be obtained from a variant of the brane configurations 
just described,
either with or without an orientifold plane.
However, $\su{N}$ gauge theory with {\it two} antisymmetric hypermultiplets
apparently requires a configuration with at least two O$6^-$ planes,
as described in Ref.~\cite{twoanti} and 
in the talk of H. J. Schnitzer at this Workshop \cite{montrealone}.
These two O$6^-$ planes, moreover, 
generate an infinite number of O$6^-$ planes and NS 5-branes, 
equally spaced in the $x_6$ direction.
The corresponding SW curve would be of infinite order.

Alternatively, 
we may observe that a pair of reflections 
through different points generates a translation, 
so the brane configuration must in fact be periodic in the $x_6$ direction.
A second compactified direction, $x_{10}$, emerges in M-theory,
so that the lifted M5 configuration lives on a torus.
We are thus naturally led to a discussion of elliptic models 
\cite{DonagiWitten,WittenM,Uranga}.
The infinite order SW curve may be regarded as an elliptic curve
written on the covering space (see, e.g., ref.~\cite{Yokono}).

We now describe the class of elliptic brane configurations,
containing a pair of O$6^{\pm}$ planes,
that give rise to four-dimensional gauge theories with
simple gauge groups and vanishing beta function \cite{Uranga}.
The latter condition requires that the total 6-brane charge vanish,
so that configurations with two O$6^-$ planes also
contain four D6 branes (plus mirrors) parallel to the O6 planes;
configurations with O$6^+$ and O$6^-$ require no D6 branes.

\begin{picture}(430,200)(10,10)

\put(30,50){\line(0,1){150}}
\put(100,50){\line(0,1){150}}
\put(170,50){\line(0,1){150}}

\put(10,190){{\bf(a)}}


\put(70,42){\vector(-1,0){40}}
\put(78,38){unit cell}
\put(130,42){\vector(1,0){40}}



\put(26,134){$\otimes$}
\put(35,134){O$6^-$}
\put(96,124){$\otimes$}
\put(105,124){O$6^-$}
\put(166,114){$\otimes$}

\put(30,80){\line(1,0){9}}
\put(49,80){\line(1,0){9}}
\put(68,80){\line(1,0){9}}
\put(87,80){\line(1,0){9}}

\put(30,100){\line(1,0){9}}
\put(49,100){\line(1,0){9}}
\put(68,100){\line(1,0){9}}
\put(87,100){\line(1,0){9}}

\put(65,120){$\cdot$}
\put(65,125){$\cdot$}
\put(65,130){$\cdot$}
\put(65,115){$\cdot$}

\put(35,156){\framebox(5,5){$\cdot$}}
\put(52,156){\framebox(5,5){$\cdot$}}
\put(69,156){\framebox(5,5){$\cdot$}}
\put(86,156){\framebox(5,5){$\cdot$}}

\put(100,168){\line(1,0){9}}
\put(119,168){\line(1,0){9}}
\put(138,168){\line(1,0){9}}
\put(157,168){\line(1,0){9}}

\put(100,148){\line(1,0){9}}
\put(119,148){\line(1,0){9}}
\put(138,148){\line(1,0){9}}
\put(157,148){\line(1,0){9}}

\put(135,120){$\cdot$}
\put(135,125){$\cdot$}
\put(135,130){$\cdot$}
\put(135,135){$\cdot$}

\put(108,88){\framebox(5,5){$\cdot$}}
\put(125,88){\framebox(5,5){$\cdot$}}
\put(142,88){\framebox(5,5){$\cdot$}}
\put(159,88){\framebox(5,5){$\cdot$}}


\put(215,120){\vector(0,1){30}}
\put(214,153){$v$}
\put(215,120){\vector(1,0){30}}
\put(249,118){$x_6$}

\put(270,190){{\bf(b)}}
\put(290,50){\line(0,1){150}}
\put(360,50){\line(0,1){150}}
\put(430,50){\line(0,1){150}}

\put(330,42){\vector(-1,0){40}}
\put(338,38){unit cell}
\put(390,42){\vector(1,0){40}}

\put(286,134){$\otimes$}
\put(295,134){O$6^+$}
\put(356,124){$\otimes$}
\put(365,124){O$6^-$}
\put(426,114){$\otimes$}

\put(290,80){\line(1,0){9}}
\put(309,80){\line(1,0){9}}
\put(328,80){\line(1,0){9}}
\put(347,80){\line(1,0){9}}

\put(290,100){\line(1,0){9}}
\put(309,100){\line(1,0){9}}
\put(328,100){\line(1,0){9}}
\put(347,100){\line(1,0){9}}

\put(325,120){$\cdot$}
\put(325,125){$\cdot$}
\put(325,130){$\cdot$}
\put(325,115){$\cdot$}

\put(360,148){\line(1,0){9}}
\put(379,148){\line(1,0){9}}
\put(398,148){\line(1,0){9}}
\put(417,148){\line(1,0){9}}

\put(360,168){\line(1,0){9}}
\put(379,168){\line(1,0){9}}
\put(398,168){\line(1,0){9}}
\put(417,168){\line(1,0){9}}

\put(395,120){$\cdot$}
\put(395,125){$\cdot$}
\put(395,130){$\cdot$}
\put(395,135){$\cdot$}

\put(220,10){\makebox(0,0)[b]{\bf {Figure 4}}}
\end{picture}

Figures 4 and 5 show only the unit cell of the periodic configurations;
the left-and right-most NS 5-branes are to be identified
(with a possible shift in the $v$ direction).
The corresponding gauge theories may be identified using
the rules described above.
The configurations in fig.~4 contain two NS 5-branes per unit cell.
Figure~4(a) corresponds to 
$\su{N}$ gauge theory with two antisymmetric hypermultiplets
and four fundamental hypermultiplets,
whose degrees of freedom arise from strings stretched between
the D4- and D6-branes.
Figure~4(b) corresponds to 
$\su{N}$ gauge theory with an antisymmetric and a symmetric hypermultiplet.

\begin{picture}(430,200)(10,10)

\put(30,50){\line(0,1){150}}
\put(170,50){\line(0,1){150}}

\put(10,190){{\bf(a)}}

\put(70,42){\vector(-1,0){40}}
\put(78,38){unit cell}
\put(130,42){\vector(1,0){40}}




\put(26,134){$\otimes$}
\put(35,134){O$6^-$}
\put(96,124){$\otimes$}
\put(105,124){O$6^-$}
\put(166,114){$\otimes$}

\put(30,80){\line(1,0){9}}
\put(49,80){\line(1,0){9}}
\put(68,80){\line(1,0){9}}
\put(87,80){\line(1,0){9}}
\put(106,80){\line(1,0){9}}
\put(125,80){\line(1,0){9}}
\put(144,80){\line(1,0){9}}
\put(163,80){\line(1,0){7}}

\put(30,100){\line(1,0){9}}
\put(49,100){\line(1,0){9}}
\put(68,100){\line(1,0){9}}
\put(87,100){\line(1,0){9}}
\put(106,100){\line(1,0){9}}
\put(125,100){\line(1,0){9}}
\put(144,100){\line(1,0){9}}
\put(163,100){\line(1,0){7}}

\put(65,120){$\cdot$}
\put(65,125){$\cdot$}
\put(65,130){$\cdot$}
\put(65,115){$\cdot$}

\put(35,156){\framebox(5,5){$\cdot$}}
\put(52,156){\framebox(5,5){$\cdot$}}
\put(69,156){\framebox(5,5){$\cdot$}}
\put(86,156){\framebox(5,5){$\cdot$}}

\put(30,168){\line(1,0){9}}
\put(49,168){\line(1,0){9}}
\put(68,168){\line(1,0){9}}
\put(87,168){\line(1,0){9}}
\put(106,168){\line(1,0){9}}
\put(125,168){\line(1,0){9}}
\put(144,168){\line(1,0){9}}
\put(163,168){\line(1,0){7}}

\put(30,148){\line(1,0){9}}
\put(49,148){\line(1,0){9}}
\put(68,148){\line(1,0){9}}
\put(87,148){\line(1,0){9}}
\put(106,148){\line(1,0){9}}
\put(125,148){\line(1,0){9}}
\put(144,148){\line(1,0){9}}
\put(163,148){\line(1,0){7}}

\put(135,120){$\cdot$}
\put(135,125){$\cdot$}
\put(135,130){$\cdot$}
\put(135,135){$\cdot$}

\put(108,88){\framebox(5,5){$\cdot$}}
\put(125,88){\framebox(5,5){$\cdot$}}
\put(142,88){\framebox(5,5){$\cdot$}}
\put(159,88){\framebox(5,5){$\cdot$}}


\put(215,120){\vector(0,1){30}}
\put(214,153){$v$}
\put(215,120){\vector(1,0){30}}
\put(249,118){$x_6$}

\put(290,50){\line(0,1){150}}
\put(430,50){\line(0,1){150}}

\put(270,190){{\bf(b)}}

\put(330,42){\vector(-1,0){40}}
\put(338,38){unit cell}
\put(390,42){\vector(1,0){40}}

\put(286,134){$\otimes$}
\put(295,134){O$6^\pm$}
\put(356,124){$\otimes$}
\put(365,124){O$6^\mp$}
\put(426,114){$\otimes$}

\put(290,80){\line(1,0){9}}
\put(309,80){\line(1,0){9}}
\put(328,80){\line(1,0){9}}
\put(347,80){\line(1,0){9}}
\put(366,80){\line(1,0){9}}
\put(385,80){\line(1,0){9}}
\put(404,80){\line(1,0){9}}
\put(423,80){\line(1,0){7}}

\put(290,100){\line(1,0){9}}
\put(309,100){\line(1,0){9}}
\put(328,100){\line(1,0){9}}
\put(347,100){\line(1,0){9}}
\put(366,100){\line(1,0){9}}
\put(385,100){\line(1,0){9}}
\put(404,100){\line(1,0){9}}
\put(423,100){\line(1,0){7}}

\put(325,120){$\cdot$}
\put(325,125){$\cdot$}
\put(325,130){$\cdot$}
\put(325,115){$\cdot$}

\put(290,148){\line(1,0){9}}
\put(309,148){\line(1,0){9}}
\put(328,148){\line(1,0){9}}
\put(347,148){\line(1,0){9}}
\put(366,148){\line(1,0){9}}
\put(385,148){\line(1,0){9}}
\put(404,148){\line(1,0){9}}
\put(423,148){\line(1,0){7}}

\put(290,168){\line(1,0){9}}
\put(309,168){\line(1,0){9}}
\put(328,168){\line(1,0){9}}
\put(347,168){\line(1,0){9}}
\put(366,168){\line(1,0){9}}
\put(385,168){\line(1,0){9}}
\put(404,168){\line(1,0){9}}
\put(423,168){\line(1,0){7}}

\put(395,120){$\cdot$}
\put(395,125){$\cdot$}
\put(395,130){$\cdot$}
\put(395,135){$\cdot$}

\put(220,10){\makebox(0,0)[b]{\bf {Figure 5}}}
\end{picture}

The configurations in fig.~5 contain only one NS 5-brane per unit cell.
Figure~5(a) corresponds to 
$\sp{2N}$ gauge theory with an antisymmetric hypermultiplet
and four fundamental hypermultiplets.
Figure~5(b) corresponds to 
$\sp{2N}$ gauge theory with an adjoint hypermultiplet
(O$6^+$ on the NS 5-brane), or
$\so{2N}$ gauge theory with an adjoint hypermultiplet
(O$6^-$ on the NS 5-brane).

Finally, a periodic configuration without orientifold planes
and with one NS 5-brane per unit cell corresponds to
$\su{N}$ gauge theory with an adjoint hypermultiplet \cite{WittenM}.
\vspace{.3in}

\noindent{\bf 3. M theory and Seiberg-Witten curves}
\renewcommand{\theequation}{3.\arabic{equation}}
\setcounter{equation}{0}

In the strong-coupling limit, type IIA string theory 
goes over to eleven-dimensional M-theory 
with an additional periodic coordinate $x_{10}$ (with period $R$);
the brane configurations described in the previous section 
are ``lifted'' to M5-brane configurations \cite{WittenM}.
The M5-brane world-volume is  $\IR^4 \times \Sigma$
where $\IR^4$ spans the 0123 directions, 
and $\Sigma$ is a two-dimensional submanifold of $Q \sim \IC^2 = (v,t)$,
where $v=x_4+i x_5$ and $ t = \exp \left[-(x_6 + i x_{10})/R\right]$.
The M5-brane is located at a point in the remaining 789 directions.
$\Sigma \subset Q$ can be written as an algebraic curve,
which is none other than the Seiberg-Witten curve of the 
corresponding four-dimensional gauge theory.

The M5-brane curve $\Sigma$ corresponding 
to the IIA configuration shown in fig.~1 is \cite{WittenM, ENR}
\be
t^3 - \prod_{i=1}^{N_1} (v-a_i) \, t^2 
+ \Lambda^{2N_1-N_2} \prod_{i=1}^{N_2} (v-b_i) \, t 
- \Lambda^{3N_1} = 0.
\label{cubic}
\ee
The features of this curve can be understood directly 
in terms of the classical IIA picture.
Holding $v$ fixed, eq.~\eqs{cubic} has three solutions for $t$;
these correspond to the positions of the NS 5-branes.
The coefficients of the various powers of $t$ 
vanish at the positions of the D4-branes between adjacent NS 5-branes.
Since there are no D4 branes to the left and right of the NS 5-branes,
the first and last terms of the curve have constant coefficients.
The curve \eqs{cubic} is indeed the SW curve of the 
$\su{N_1} \times \su{N_2}$ gauge theory associated with this IIA configuration.

The M-theory geometry corresponding to a type IIA configuration
involving an orientifold plane is more complicated.
For a single O$6^-$ plane, as in fig.~2,
the M-theory background is an Atiyah-Hitchin space \cite{AH}.
This may be described in terms of 
a submanifold $\tilde Q$ of $\IC^3 = (v, t_L, t_R)$.
Far from the orientifold plane, $\tilde Q$ is given by \cite{LandsteinerLopez}
\be
t_L t_R^{-1} = {\Lambda^{2N+4} \over \left(v + \half m \right)^4}
\label{atiyah}
\ee
and is invariant under
\be
v \to -v-m, \qquad  t_L \to t_R^{-1}.
\label{invariance}
\ee
In the region far to the left of the orientifold plane
($x_6 \to -\infty$),
the variable $t_L \to t$,
whereas in the region far to the right of the orientifold plane
($x_6 \to \infty$),
the variable $t_R \to t$.

The M5-brane configuration corresponding to the type IIA configuration
shown in fig.~2 is given by $\IR^4 \times \Sigma$,
where now $\Sigma$ is an algebraic curve embedded in $\tilde Q$,
\be
t_L^3 + \prodN (v- a_i) \, t_L^2 + A(v) \, t_L + B(v) = 0.
\label{Lcurve}
\ee
Since $t_L$ corresponds to $t$ to the left of the orientifold plane,
the coefficients of the first two (but not the last two) terms
correspond to the positions of the D4-branes in that region of fig.~2.
The curve $\Sigma$ must be invariant under \eqs{invariance},
and so may be rewritten as
\be
B(-v-m) \, t_R^3 + A(-v-m) \, t_R^2 + \prodN (-v-m-a_i) \, t_R + 1 = 0
\label{Rcurve}
\ee
where now the coefficients of the last two terms correspond
to the positions of the D4-branes to the right of the orientifold plane
in fig.~2.
Using eq.~\eqs{atiyah}, we may rewrite eq.~\eqs{Rcurve} in terms of $t_L$;
equating the result with eq.~\eqs{Lcurve}, we finally obtain
\be
t_L^3 + \prodN (v- a_i)\, t_L^2 + 
{\Lambda^{N+2} \prodN (-v-a_i-m) \over (v+\half m)^2} \, t_L + 
{\Lambda^{3N+6}  \over (v+\half m)^6} = 0.
\label{oricurve}
\ee
We have used eq.~\eqs{atiyah}, 
which is valid only far from the orientifold;
consequently, eq. \eqs{oricurve} only gives the leading terms 
(in powers of $\Lambda$) of the curve.
The subleading terms may be determined by a more careful
consideration of the Atiyah-Hitchin space $\tilde Q$ \cite{LandsteinerLopez}.
Including the subleading terms, and defining
$y = t_L/ (v+\half m)^2$,
we obtain the curve
\be
y^3 &+& y^2 \left[ (v+\half m)^2 \prodN (v-a_i)+ 3 \Lambda^{N+2} \right]\nn\\
&+& y \Lambda^{N+2} 
\left[(v+\half m)^2 \prodN (-v-a_i-m) + 3 \Lambda^{N+2} \right]
~+~  \Lambda^{3N+6} ~=~ 0,
\ee
which is the form of the curve given by 
Landsteiner and Lopez \cite{LandsteinerLopez}.
{}From this curve, the one-instanton propotential 
may be calculated \cite{oneanti}.

After these warm-ups, we turn to the calculation 
of the SW curves for the elliptic IIA configurations
shown in figs.~4 and 5.
For concreteness, we focus on the configuration in fig.~4(b),
which gives rise to the four-dimensional ${\cal N}=2$ supersymmetric
$\su{N}$ gauge theory with hypermultiplets in the symmetric $\Ysymm$ 
and antisymmetric $\Yasymm$ representations, 
but the other cases are analogous.

\begin{center}
\begin{picture}(810,295)(10,10)


\put(10,192){\line(1,0){5}}
\put(20,192){\line(1,0){5}}
\put(30,192){\line(1,0){5}}
\put(40,192){\line(1,0){5}}
\put(50,192){\line(1,0){5}}
\put(60,192){\line(1,0){5}}
\put(70,192){\line(1,0){5}}
\put(80,192){\line(1,0){5}}
\put(90,192){\line(1,0){5}}
\put(100,192){\line(1,0){5}}
\put(5,180){$(\!v\!-\!a_i\!+\!m_2\!-\!m_1\!)$}

\put(105,240){\line(1,0){2}}
\put(110,240){\line(1,0){5}}
\put(120,240){\line(1,0){5}}
\put(130,240){\line(1,0){5}}
\put(140,240){\line(1,0){5}}
\put(150,240){\line(1,0){5}}
\put(160,240){\line(1,0){5}}
\put(170,240){\line(1,0){5}}
\put(180,240){\line(1,0){5}}
\put(190,240){\line(1,0){5}}
\put(200,240){\line(1,0){5}}
\put(120,245){$(\!v\!+\!a_i\!+\!m_2\!)$}

\put(50,-10){$t_{2}$}
\put(150,-10){$t_{1}$}
\put(250,-10){$t_0$}
\put(350,-10){$t_{-1}$}
\put(450,-10){$t_{-2}$}

\put(205,117){\line(1,0){2}}
\put(210,117){\line(1,0){5}}
\put(220,117){\line(1,0){5}}
\put(230,117){\line(1,0){5}}
\put(240,117){\line(1,0){5}}
\put(250,117){\line(1,0){5}}
\put(260,117){\line(1,0){5}}
\put(270,117){\line(1,0){5}}
\put(280,117){\line(1,0){5}}
\put(290,117){\line(1,0){5}}
\put(300,117){\line(1,0){5}}
\put(235,105){$(\!v-\!a_i\!)$}

\put(305,155){\line(1,0){2}}
\put(310,155){\line(1,0){5}}
\put(320,155){\line(1,0){5}}
\put(330,155){\line(1,0){5}}
\put(340,155){\line(1,0){5}}
\put(350,155){\line(1,0){5}}
\put(360,155){\line(1,0){5}}
\put(370,155){\line(1,0){5}}
\put(380,155){\line(1,0){5}}
\put(390,155){\line(1,0){5}}
\put(400,155){\line(1,0){5}}
\put(320,160){$(\!v\!+\!a_i\!+\!m_1\!)$}

\put(405,10){\line(1,0){2}}
\put(410,10){\line(1,0){5}}
\put(420,10){\line(1,0){5}}
\put(430,10){\line(1,0){5}}
\put(440,10){\line(1,0){5}}
\put(450,10){\line(1,0){5}}
\put(460,10){\line(1,0){5}}
\put(470,10){\line(1,0){5}}
\put(480,10){\line(1,0){5}}
\put(407,15){$(\!v\!-\!a_i\!+\!m_1\!-\!m_2\!)$}


\put(105,5){\line(0,1){255}}
\put(205,5){\line(0,1){255}}
\put(305,5){\line(0,1){255}}
\put(405,5){\line(0,1){255}}


\put(101,215){$\otimes$} 
\put(201,175){$\otimes$}
\put(301,135){$\otimes$}
\put(401,85){$\otimes$}
\put(31,220){$(\!v\!+\!m_2\!-\!\half m_1\!)$}
\put(160,170){$(\!v\!+\!\half m_2\!)$}
\put(259,135){$(\!v\!+\!\half m_1\!)$}
\put(406,75){$(\!v\!+\!m_1\!-\!\half m_2\!)$}

\put(110,220){O$6^{-}$}
\put(210,180){O$6^{+}$}
\put(310,140){O$6^{-}$}
\put(410,90){O$6^{+}$}

\put(30,30){\vector(1,0){20}}
\put(30,30){\vector(0,1){20}}
\put(25,50){$v$}
\put(50,25){$x_6$}

\put(220,-40){\makebox(0,0)[b]{\bf {Figure 6}}}
\end{picture}
\end{center}

\vs{.6in}
Figure 6 shows a piece of the IIA configuration on the covering space.
The displacements of the O6 planes in the $v$ direction
correspond to the masses of the hypermultiplets;
here $m_1$ ($m_2$) is the mass 
of the antisymmetric (symmetric) hypermultiplet.

We would like to derive the M5-brane curve $\IR^4 \times \Sigma$
that corresponds to the configuration in fig.~6.
First we must describe the M-theory geometry 
in which $\Sigma$ is embedded.
Because of the presence of orientifold planes, 
we introduce an infinite set of ``local'' variables $t_n$.
In the region between any pair of orientifold planes, 
the corresponding variable $t_n$ shown in fig.~6 
corresponds to $ t = \exp \left[-(x_6 + i x_{10})/R\right]$.
The two variables $t_0$ and $t_{-1}$ adjacent to 
the O$6^-$ plane at $v=-\half m_1$
are related, far from the plane, by
\be
t_0 t_{-1}^{-1} = {q^{1/2} \over \left(v + \half m_1 \right)^4}
\label{transminus}
\ee
as in eq.~\eqs{atiyah}, 
where for elliptic models the 
parameter $q$ replaces
the scale factor $\Lambda$.
The two variables $t_0$ and $t_1$ adjacent to the O$6^+$ plane 
at $v=-\half m_2$
are related, far from the plane, by \cite{LandsteinerLopez}
\be
t_1 t_{0}^{-1} = q^{1/2} \left(v + \half m_2 \right)^4.
\label{transplus}
\ee
The pairs of variables adjacent to each of the other O6 planes 
are related by analogous ``transition functions.''

Let us write the (leading terms of the) infinite order
curve $\Sigma$ as 
\be
\cdots 
~+~ q P_2 (v) t_0^2
~+~ q^{1/4} P_1 (v) t_0
~+~ P_0 (v) 
~+~ q^{1/4} P_{-1} (v) t_0^{-1}
~+~ q P_{-2} (v) t_0^{-2} ~+~ \cdots~=~0
~~~~~~\label{firstcurve}
\ee
in terms of the local variable $t_0$.
In the central region of fig.~6,
$t_0$ corresponds to $t$,
so the coefficient $P_0(v)$ must vanish at the positions
$v = a_i$ of the $N$ D4 branes in that region,
i.e. $P_0 (v) = \prodN (v-a_i)$.
To determine the other coefficients, $P_n(v)$,  
we must use the invariance of the curve induced by the
orientifold planes, 
together with the ``transition functions'' 
\eqs{transminus} and \eqs{transplus}.
First, requiring that the curve be invariant under
\be
v \to -v-m_1, \qquad  t_0 \to t_{-1}^{-1}
\label{firstinvariance}
\ee
we may rewrite eq.~\eqs{firstcurve}  as
\be
&\cdots& 
 +~ q^{1/4} P_1 (-v-m_1) \,t_{-1}^{-1}
~+~ P_0 (-v-m_1) \\
&&+~ q^{1/4} P_{-1} (-v-m_1) \,t_{-1}
~+~ q P_{-2} (-v-m_1)\, t_{-1}^2 + \cdots=0.\nn
\ee
We then use eq.~\eqs{transminus} to obtain
\be
&\cdots &
 +~ q^{1/4} (v+\half m_1)^6 P_{-2} (-v-m_1) t_0
~+~         (v+\half m_1)^2 P_{-1} (-v-m_1)   \\
&&+~ q^{1/4} (v+\half m_1)^{-2} P_0 (-v-m_1) t_0^{-1}
~+~ q       (v+\half m_1)^{-6} P_1 (-v-m_1) t_0^{-2} + \cdots = 0 \nn
\ee
Equating this with eq.~\eqs{firstcurve}, we find
\be
P_{-1}(v) &=&  (v+\half m_1)^{-2} P_0 (-v-m_1), \nn\\
P_{-2}(v) &=&  (v+\half m_1)^{-6} P_1 (-v-m_1), \\
	&\vdots &\nn
\la{relations}
\ee
Next, we require that the curve be invariant under
\be
v \to -v-m_2, \qquad  t_0 \to t_{1}^{-1}.
\label{secondinvariance}
\ee
This, together with eq.~\eqs{transplus}, 
generates a set of relations similar to eq.~\eqs{relations}.
The two reflections \eqs{firstinvariance} and \eqs{secondinvariance}
generate the entire invariance group,
and so are sufficient to give us
(the leading terms of) the entire set of coefficients
\be 
	&\vdots& \nn\\
P_{ 2} (v)  &=& (v+\half m_2)^{ 6} \,(v+m_2-\half m_1)^{-2} \,
                \prodN(v-a_i+m_2-m_1),  \nn\\
P_{ 1} (v)  &=& (v+\half m_2)^{ 2} \,
            \prodN(-v-a_i-m_2), \nn \\
P_{ 0} (v)  &=& \prodN (v-a_i), \\
P_{-1} (v)  &=& (v+\half m_1)^{-2} \,
            \prodN (-v-a_i- m_1),  \nn \\
P_{-2} (v)  &=& (v+\half m_1)^{-6} \,(v+m_1-\half m_2)^{2} \,
                \prodN (v-a_i+m_1-m_2), \nn\\
	&\vdots& \nn
\label{coeffs}
\ee
This is equivalent to  the curve given in ref.~\cite{elliptic},
up to overall multiplication by $F(v)$
and redefinition $t_0 = G(v) t$, where 
$F(v)$ and $G(v)$ are rational functions 
of $v$ and the hypermultiplet masses.
The prepotential derived from these curves is independent of $F(v)$ and $G(v)$.

In Refs.~\cite{twoanti,elliptic}, SW curves are given for all 
the other elliptic models discussed in the previous section.

In the limit $m_1 \to m_2$ (i.e., vanishing global mass),
there are no subleading terms, and the curve \eqs{firstcurve}
for $\su{N} + \Ysymm + \Yasymm$ reduces,
upon change of variable $t = t_0 (v+ \half m)^2$,
to 
\be
0 &=&
\sum_{n\,{\rm even}}\, q^{n^2/4} \, t^n\,\prodN (v-a_i)\,+
\sum_{n\,{\rm odd}} \, q^{n^2/4} \, t^n\,\prod_{i=1}^{N}(-v-a_i-m)   \nn \\
&=&
\theta_3 \left({z \over  \omega_1} | 2\tau\right) \, \prodN(v-a_i)\,+
\theta_2 \left({z \over  \omega_1} | 2\tau\right)
 \,\prodN(-v-a_i- m),
\la{ellipcurve}
\ee
where $q=\exp(2\pi i \tau)$, $t = \exp (-i\pi z/\omega_1)$,
and $\theta_2(\nu|\tau)$, $\theta_3(\nu|\tau)$  are Jacobi theta functions.
In ref.~\cite{elliptic}, we have shown that eq.~\eqs{ellipcurve} 
is equivalent to the curve for this theory given by Uranga \cite{Uranga}.

\np
\noindent{\bf 4. One-instanton prepotential}
\renewcommand{\theequation}{4.\arabic{equation}}
\setcounter{equation}{0}

Although we have obtained an infinite order curve for the
$\su{N} + \Ysymm + \Yasymm$ theory, 
the one-instanton $({\cal O}(q))$ prepotential for this theory may
be extracted \cite{product,twoanti} from the quartic truncation 
of this curve consisting of just those five terms shown explicitly in eq.~\eqs{coeffs}.
Define 
\be
S(v) = {P_1(v) P_{-1}(v) \over P_0(v)^2}.
\ee
For this theory, $S(v)$ has quadratic poles at $v=a_k$ and $v=-\half m_1$.
At these poles, we define the residue functions $S_k(v)$ and $S_{m_1}(v)$ by
\be
S(v)\,=\,{S_k(v)\over (v-a_k)^2}\, =\,{S_{m_1} (v)\over (v+\half {m_1} )^2}.
\label{residue} 
\ee 
It may be shown that the one-instanton prepotential is given by
\be
2 \pi i {\cal F}_{\rm 1-inst} = \, 
\sum_{k=1}^N S_k(a_k) -2S_{m_1}(-\half m_1).
\label{oneinst}
\ee
Although $S(v)$ and therefore eq.~\eqs{oneinst}
depend explicitly only on three of the five coefficients 
in eq.~\eqs{coeffs},
the entire quartic truncation 
(including the first subleading term) 
is necessary for the consistency
of the calculation to ${\cal O}(q)$ \cite{oneanti}.

In Table 1 below, 
we list the expressions for $S(v)$ 
for all the elliptic models described in sect. 2 \cite{elliptic}.
The one-instanton prepotential for each of these theories is then
given in terms of the residue functions defined in eq.~\eqs{residue}.
For $\su{N} + $ adjoint \cite{DHokerPhong}
\be
 2\pi i {\cal F}_{\rm 1-inst} =  
\sum_{k=1}^N S_k(a_k).
\ee
For $\su{N} + \Yasymm + \Ysymm$,
$\so{2N} + $ adjoint, and
$\so{2N+1} + $ adjoint, 
\be
 2\pi i {\cal F}_{\rm 1-inst} =  
\sum_{k=1}^N S_k(a_k) -2S_{m_1}(-\half m_1),
\ee
where $m_1$ is the mass of the antisymmetric or adjoint hypermultiplet.
For 
$ \su{N}  +2 \,\Yasymm + 4\,\fund  $,
\be
 2\pi i {\cal F}_{\rm 1-inst} =  
\sum_{k=1}^N S_k(a_k)-2S_{m_1}(-\half m_1)-2S_{m_2}(-\half m_2),
\ee
where $m_1$ and $m_2$ are the masses of the antisymmetric hypermultiplets.
For  $\sp{2N} + $ adjoint, and 
$\sp{2N} + \Yasymm + 4\,\fund$, 
\be
 2\pi i {\cal F}_{\rm 1-inst} =  
-2 [\quarS_{0}(0)]^{1/2}
\ee
where 
\be
S(v)\,=\, {\quarS_0(v) \over v^4}
\label{quartic} 
\ee 
defines the residue function at the quartic pole at $v=0$.
All of these results have been subjected to a wide variety of
consistency checks, as described in ref.~\cite{elliptic}.

\begin{center} 
\begin{tabular}{||c|c||} 
\hline
\hline ${\cal N}=2$ gauge theory & $S(v)$ \\
\hline\hline  
$ \su{N} \, +\,2 \,\Yasymm\,(m_1,m_2) \,+\, 4\,\fund \,(M_j)  $ 
& 
${{ \prod_{i=1}^{N}(v+a_i+m_1) \prod_{i=1}^N(v +a_i+m_2) 
\prod_{j=1}^{4}(v +M_j)
\vphantom{\bigg|} \over \vphantom{\bigg|} 
(v +\half m_1)^2 (v +\half m_2)^2
\prod_{i=1}^N (v -a_i)^2}}$ 
\\ [.1in]
\hline 
$ \su{N} +   \Yasymm (m_1) +  \Ysymm (m_2) $ 
& 
${(v+\half m_2)^2\,
\prod_{i=1}^N (v+a_i+m_1)  \prod_{i=1}^N (v+a_i+m_2) 
\vphantom{\bigg|} \over \vphantom{\bigg|} 
(v+\half m_1)^2 \prod_{i=1}^N (v-a_i)^2}$ 
\\ [.1in]
\hline  
$\su{N}  + {\rm adjoint} (m) $
& 
${\prod_{i=1}^N [(v-a_i)^2-m^2] 
\vphantom{\bigg|} \over \vphantom{\bigg|} 
\prod_{i=1}^N (v-a_i)^2}$ 
\\ [.1in]
\hline 
$\so{2N}+ {\rm adjoint} (m) $ 
& 
$ {v^4 \prod_{i=1}^N [(v-m)^2 -a_i^2] \prod_{i=1}^N [(v+m)^2 -a_i^2]
\vphantom{\bigg|} \over \vphantom{\bigg|} 
{ (v+\ha m)^2 (v-\ha m)^2 \prod_{i=1}^N (v^2-a_i^2)^2}}$       
\\ [.1in]
\hline 
$\so{2N+1} + {\rm adjoint} (m)$ 
& 
$ {v^2 (v+m)(v-m) \prod_{i=1}^N [(v-m)^2 -a_i^2] 
\prod_{i=1}^N [(v+m)^2 -a_i^2]
\vphantom{\bigg|} \over \vphantom{\bigg|} 
{ (v+\ha m)^2 (v-\ha m)^2 \prod_{i=1}^N (v^2-a_i^2)^2}}$       
\\ [.1in]
\hline 
$\sp{2N}+ {\rm adjoint} (m)$
& 
$ {(v+\ha m)^2 (v-\ha m)^2 \prod_{i=1}^N [(v-m)^2 -a_i^2]
\prod_{i=1}^N [(v+m)^2 -a_i^2]
\vphantom{\bigg|} \over \vphantom{\bigg|} 
{ v^4\prod_{i=1}^N (v^2-a_i^2)^2}}$       
\\ [.1in]
\hline 
$\sp{2N}\, + \,\Yasymm (m) \,+\,4\, \fund (M_j)$ 
& 
$ {\prod_{i=1}^N [(v-m)^2 -a_i^2] \prod_{i=1}^N [(v+m)^2 -a_i^2]
\prod_{j=1}^{4}(v^2-M_j^2)
\vphantom{\bigg|} \over \vphantom{\bigg|} 
{ v^4 (v+\ha m)^2 (v-\ha m)^2  \prod_{i=1}^N (v^2-a_i^2)^2}}$       
\\ [.1in]
\hline 
\hline 
\end{tabular} 
\end{center} 
\label{tablebis}
\centerline{\footnotesize{\bf Table 1}}
\vspace{0.2cm}

\centerline{\bf Acknowledgement}

We are grateful to the organizers of this Workshop,
E. D'Hoker, D.H.  Phong, and S.T. Yau,
for the opportunity to present this work
in such a pleasant and stimulating environment.

\end{document}